\documentclass[pdflatex,sn-mathphys-num]{sn-jnl}

\usepackage[linesnumbered,ruled,vlined]{algorithm2e}
\usepackage{graphicx}%
\usepackage{multirow}%
\usepackage{amsmath,amssymb,amsfonts}%
\usepackage{amsthm}%
\usepackage{mathrsfs}%
\usepackage[title]{appendix}%
\usepackage{xcolor}%
\usepackage{textcomp}%
\usepackage{manyfoot}%
\usepackage{booktabs}%
\usepackage{algpseudocode}%
\usepackage{listings}%
\usepackage{booktabs}

\theoremstyle{thmstyleone}%
\theoremstyle{thmstyletwo}%

\theoremstyle{thmstylethree}%

\begin{document}

\title[Real-Time Forecasting of Pathological Gait via IMU Navigation: A Few-Shot and Generative Learning Framework for Wearable Devices]{Real-Time Forecasting of Pathological Gait via IMU Navigation: A Few-Shot and Generative Learning Framework for Wearable Devices}


\author*[1]{\fnm{Wenwen} \sur{Zhang}}\email{wenwenzhang@ece.ubc.ca}

\author*[1,2]{\fnm{Hao} \sur{Zhang}}\email{haozh@g.ucla.edu}

\author[1]{\fnm{Zenan} \sur{Jiang}}\email{jiang@ece.ubc.ca}

\author[1]{\fnm{Amir} \sur{Servati}}\email{amir.servati@ubc.ca}

\author[1,3]{\fnm{Peyman} \sur{Servati}}\email{peymans@ece.ubc.ca}

\affil*[1]{\orgdiv{Department of Electrical and Computer Engineering}, \orgname{University of British Columbia}, \orgaddress{\street{2332 Main Mall}, \city{Vancouver}, \postcode{V6T 1Z4}, \state{BC}, \country{Canada}}}

\affil[2]{\orgdiv{Department of Electrical and Computer Engineering}, \orgname{University of California, Los Angeles}, \orgaddress{\street{420 Westwook Plaza}, \city{Los Angeles}, \postcode{90095}, \state{CA}, \country{United States}}}


\abstract{Current gait analysis faces challenges in various aspects, including limited and poorly labeled data within existing wearable electronics databases, difficulties in collecting patient data due to privacy concerns, and the inadequacy of the Zero-Velocity Update Technique (ZUPT) in accurately analyzing pathological gait patterns. To address these limitations, we introduce GaitMotion, a novel machine-learning framework that employs few-shot learning on a multitask dataset collected via wearable IMU sensors for real-time pathological gait analysis. GaitMotion enhances data quality through detailed, ground-truth-labeled sequences and achieves accurate step and stride segmentation and stride length estimation, which are essential for diagnosing neurological disorders. We incorporate a generative augmentation component, which synthesizes rare or underrepresented pathological gait patterns. GaitMotion achieves a 65\% increase in stride length estimation accuracy compared to ZUPT. In addition, its application to real patient datasets via transfer learning confirms its robust predictive capability. By integrating generative AI into wearable gait analysis, GaitMotion not only refines the precision of pathological gait forecasting but also demonstrates a scalable framework for leveraging synthetic data in biomechanical pattern recognition, paving the way for more personalized and data-efficient digital health services.}




\maketitle

\section{Introduction}

Gait analysis \cite{chen2016toward, 9466394}, recognized for its complexity and uniqueness akin to fingerprints, serves as a crucial tool across diverse fields, including medicine, security, rehabilitation, and sports science \cite{lonini2018wearable, czech2020age}. It is not only instrumental in personal identification through unique gait patterns \cite{zhu2021gait,  huang20213d}, but also plays a pivotal role in medical applications. The in-depth study of the gait cycle and its parameters is vital for diagnosing and treating various neurological disorders \cite{ jiang2024short,10353997} such as Parkinson's, Huntington's, and Stroke, along with orthopedic issues \cite{webster2019principles, ozerlat2012interactive} like skeletal deformities and injuries, and other medical conditions including arthritis and deep vein thrombosis. This knowledge is further applied in developing medical technologies like musculoskeletal models \cite{weng2021natural} and prosthetics \cite{kidzinski2020artificial}. These technologies, aiding in generating effective and physically plausible walking tasks \cite{takeishi2021variational}, \cite{park2022generative}, not only demonstrate the significant potential of gait analysis in enhancing the quality of life for individuals with mobility issues but also pave the way for the integration of advanced tools in clinical settings. Among these tools, optical cameras equipped with 3D markers \cite{kressig2004temporal} have emerged as a primary method for assisting with the measurement of temporal and spatial gait parameters in a clinical context. While optical systems offer precision in gait and movement analysis, their efficacy is often compromised in dynamic, real-world settings due to challenges such as occlusions and varying lighting conditions. These systems typically require controlled lab environments, making them unsuitable for continuous, long-term daily activity monitoring of patients. 

\paragraph{Wearable Sensors and Gait Analysis} 
The GaitRite system is favored as the gold standard in clinical settings over optical systems due to its ease of use, real-world applicability, non-intrusiveness, cost-effectiveness, and data accuracy \cite{bilney2003concurrent, menz2004reliability, albert2020evaluation}. However, it is constrained by its fixed length, susceptibility to data inaccuracies if not properly laid flat, and inability to capture intricate 3D movements\cite{bilney2003concurrent, menz2004reliability, mcdonough2001validity}. Additionally, GaitRite faces challenges in accurately recognizing steps and calculating gait parameters when confronted with freezing gaits, adhesive footsteps, and the use of walking aids.
To circumvent those issues and obtain reliable gait parameters, wearable technologies, and inertial measurement sensors \cite{klucken2013unbiased} are utilized for unobtrusive tracking of daily activities. 

For the gait analysis focusing on the wearable sensor-based approach and datasets, numerous daily scenarios including trajectory estimation \cite{Pyshoe}, gait-based activity recognition \cite{martindale2018smart}, and foot–ankle kinematic modeling \cite{davarzani2023closing} are present. Healthy participants are usually the main subjects in most gait databases. The lack of evaluation of patients with neurological disorders may be caused by the complex requirement of the comprehensive approach that includes a thorough medical history, physical examination, and diagnostic testing. Tracking their daily activities is an efficient way to assess their cognitive, motor, sensory, and autonomic functions. For example, stroke patients display typical asymmetry characteristics representing their important ability to control muscle strength, range of motion, and coordination. Parkinson's patients confront increased falling risk \cite{ullrich2022fall} due to the motor symptoms associated with the disease, such as tremors, stiffness, and impaired balance and coordination. These symptoms can affect a person's ability to walk and perform daily activities, making falls more likely. A complete dataset from wearable electronics with comprehensive labels on the step segmentation and gait parameters is beneficial for healthcare and clinical applications and can be used for multiple tasks such as classifications or estimations. 
Popular open-source datasets that focus on the gait event estimation and step segmentation are presented, including GEDS \cite{miraldo2020open}, MAREA \cite{khandelwal2017evaluation}, and OSHWSP \cite{llamas2016open}. These datasets domenstrate step segmentation and gait cycle estimations, which point to Heel Strike (HS) and Toe Off (TO) events \cite{kharb2011review}. However, except for the gait events, temporal and spatial gait parameters \cite{danion2003stride} are also critical indices that can infer the severity of gait disturbance. The subsequent dataset which considers gait parameters emerges: eGait \cite{rampp2014inertial, barth2015stride}. eGait primarily concentrates on the walking patterns of Parkinson's and geriatric patients, whereas asymmetrical gait characteristics, such as those seen in stroke patients are missing. Furthermore, eGait does not include crucial gait parameters such as gait cycle and stride velocity data in its ground truth data and its primary emphasis remains on stride segmentation.

\paragraph{Gait Analysis with Few-shot Machine Learning}
The current landscape of machine learning methods in gait analysis has predominantly leaned towards vision-based datasets, primarily due to the scarcity of publicly available datasets involving wearable sensors \cite{endo2022gaitforemer,lu2021quantifying}. This reliance on vision-based systems, while extensive, highlights a notable gap in the realm of sensor-based algorithmic approaches.
In contrast, traditional methods like Zero Velocity Update (ZUPT) \cite{skog2010evaluation, skog2010zero, park2010zero} have been extensively utilized, especially in scenarios where sensor-based data is critical, such as in correcting errors in inertial navigation systems during stationary foot phases. However, its effectiveness diminishes with pathological gaits, where standard gait patterns are not observed. Additionally, ZUPT struggles with high errors in double-time integration and gravity subtraction, limiting its precision.
To overcome the constraints inherent in ZUPT methodologies, particularly their diminished efficacy in pathological gait scenarios and issues with high errors in double-time integration and gravity subtraction, advanced sensor-based objective mobility data about gait analysis is becoming prevalent to assess health risk factors, assist with clinical diagnostics, and monitor medical treatments \cite{ moon2020classification, turner2019classification}.  

In 2019, Turner et al.\cite{turner2019classification} utilized wearable sensors to classify artificially induced gait alterations. Similarly, Moon et al.~\cite{moon2020classification} employed ML techniques, using IMUs located at the ankles, to differentiate between Parkinson’s disease and essential tremor. While platforms like GEDS \cite{miraldo2020open}, MAREA \cite{khandelwal2017evaluation}, and OSHWSP \cite{llamas2016open} have provided algorithms for gait event detection using wearable sensors, the scope of these algorithms in predicting detailed gait parameters remains limited. Building on this foundation is essential; we need to advance beyond simple classification and event detection. The application of few-shot ML approach~\cite{moon2020multimodal} has become increasingly important, offering a promising solution to overcome the challenges posed by the limited availability of extensive labeled datasets in wearable sensor-based gait analysis. Precisely predicting gait parameters is a crucial step not only for supporting clinical diagnoses but also for progressing the field of digital healthcare.

In this study, we introduce GaitMotion, a novel few-shot machine learning-based framework leveraging a multitask gait dataset derived from wearable sensors, synchronized with labels from the Gaitrite system. GaitMotion is designed to record gait parameters and step segmentation, catering to both normal and abnormal walking patterns. This capability is crucial in clinical and digital healthcare contexts, enabling the differentiation of step-to-step variances in gait patterns, a key aspect in assessing gait disorders. To further enrich the diversity of training data and compensate for the scarcity of pathological gait examples, we incorporate generative AI-based data augmentation, which synthetically simulates rare gait abnormalities. GaitMotion's rich dataset, combined with both real and synthetic ground-truth-labeled sequences, supports in-depth analysis of stride-to-stride fluctuations in diverse walking types.

By deploying GaitMotion for these complex pattern analyses, we enhance the precision of gait analysis and broaden the application of computational methods in interpreting biomechanical data. We employed machine learning techniques to effectively mitigate error accumulation during the double integration process from acceleration to distance. To demonstrate the efficacy of the GaitMotion approach, we offer insights into inferring pathological gait parameters from healthy subjects, utilizing transfer learning to compensate for the limited availability of patient annotations. Compared to ZUPT, our methods demonstrate significantly enhanced performance, with a 65\% improvement in overall accuracy for pathological gait patterns. This is also notable in patients exhibiting shuffling and asymmetrical gait patterns. This approach can be adapted to data collected from various individuals, accommodating behavioral differences while significantly reducing the reliance on sensitive patient data in research institutes or laboratory environments. Moreover, it holds the potential to deliver personalized services tailored to patients exhibiting rare and infrequently recorded pathological gait patterns, utilizing non-patient-specific datasets.

\section{Methods}

\subsection{ML Framework Implementation} 
Our implementation of the model for estimating stride length incorporates a Convolutional Neural Network (CNN). This model is structured with three convolution layers, augmented by the integration of Leaky ReLU activation functions and batch normalization techniques. We train the estimation model with a learning rate of $5e^{-5}$ and a batch size of 64. The $1^{st}$ Conv2D layer has a kernel size of 2 $\times$ 1 while the $2^{nd}$ and $3^{rd}$ Conv2D layer use 5 $\times$ 1 kernel. Batch normalization and Leaky Relu are followed by the Conv2D layer. We choose softplus as the activation function after the FC layer. The whole framework is trained, validated, and test on NVIDIA GeForce RTX 4060 and AMD Ryzen 9 7845HX - 16G RAM.

\subsection{GaitMotion Dataset}

The GaitMotion dataset inllustrated in \autoref{fig:setup} is composed of:
\begin{itemize}
\item Synchronized accelerometer and gyroscope data under three different walking patterns: Normal, Parkinson's and Stroke. 
\item Step segmentation results synchronized with accelerometer and gyroscope data.
\item Dedicated gait parameters include
gait cycle (swing/stance time, double/single support time, etc), step/stride length/time.
\end{itemize}


\subsection{Data Acquisition} 
Lower body locomotion was recorded for a total of 10 subjects (six females and four males) performing different walking tasks on the Gaitrite carpet. All participants were healthy subjects with an average age of 24.8$\pm$ 3.5 years, height of 170$\pm$10 cm, and mass of 68$\pm$12 kg, without any history of musculoskeletal injuries in the past year that would affect normal walking. Each participant was asked to walk in three different patterns: (1). Normal gait trial: participants walked at comfortable self-selected speeds to the end of the Gaitrite carpet. (2). Parkinson's gait trial: participants mimicked gait impairment symptoms of Parkinson's like shuffling steps and increased cadence \cite{mirelman2019gait} according to the research results and attached video. For this reason, we utilize the term shuffling gaits to denote Parkinson's gaits. 
(3). Stroke gait trial: participants mimicked the asymmetry in gait \cite{ olney1996hemiparetic} during walking by referring to the study results of actual Stroke patients.  
Since mimicking asymmetry walking patterns like Stroke gaits might temporarily influence healthy individuals' motor patterns, we avoid walking in any specific pathological gaits for a long time. Each pattern was repeated six times in a randomized order to guarantee data authenticity. Participants were asked to stay stationary for a couple of seconds before and after the trials for calibration. Acceleration and gyroscope sometimes suffer from mean value drifts, and those stationary periods can provide a reference for calibration in later signal processing.

\subsection{Data augmentation via generative model}

To mitigate the inherent limitations in collecting sufficient pathological gait data, particularly due to privacy concerns and the scarcity of rare gait types, we adopted a published generative model~\cite{yoon2019time} for synthetic data augmentation. This approach helps to enrich the training dataset with gait patterns without the need for additional real-world recordings.

We adopt a pretrained time-series generative adversarial network to generate synthetic IMU sequences that simulate common pathological gait anomalies, including asymmetric swing, foot drop, and delayed toe-off. The model is conditioned on temporal dynamics observed in healthy subjects mimicking patient gait, and generates realistic variations in gait cycles that preserve biomechanical plausibility. A total of 50 synthetic gait sequences were generated and incorporated into the training set used in GaitMotion. These synthetic samples contributed to better generalization of the model, especially in cases where patient-like patterns were underrepresented.

\subsection{Transfer Learning}
\label{tlearning}
We present the statistical results of transfer learning results on eGait (collected from 101 subjects aged 82.1 ± 6.5 years) in \autoref{tbl:transfer}. 
From our observations, even with minimal data from actual patients coupled with labels, we can still achieve a good performance. On average, the predictions deviate from the ground truth by approximately 0.159 m. 
The essence of our dataset is clear: we strive to offer a pre-trained model tailored for patients with gait disorders. With our pre-trained model, only limited labeled data from real patients is required for fine-tuning. This is instrumental in assisting clinical practitioners in determining gait parameters for diagnosis. We previously highlighted the shortcomings of the Gaitrite systems: they aren't particularly accommodating for patients with severe gait disorders, making it challenging to obtain adequate labels. However, our dataset bridges this gap, enabling consistent monitoring of these patients with accurate estimations. After training on our dataset, minimal labels are needed to fine-tune the parameters. Consequently, we can accurately gauge patients’ gait parameters without relying on the massive patient data from the Gaitrite carpet.

\begin{figure*}[ht]
\centering
{\includegraphics[width=1\linewidth]{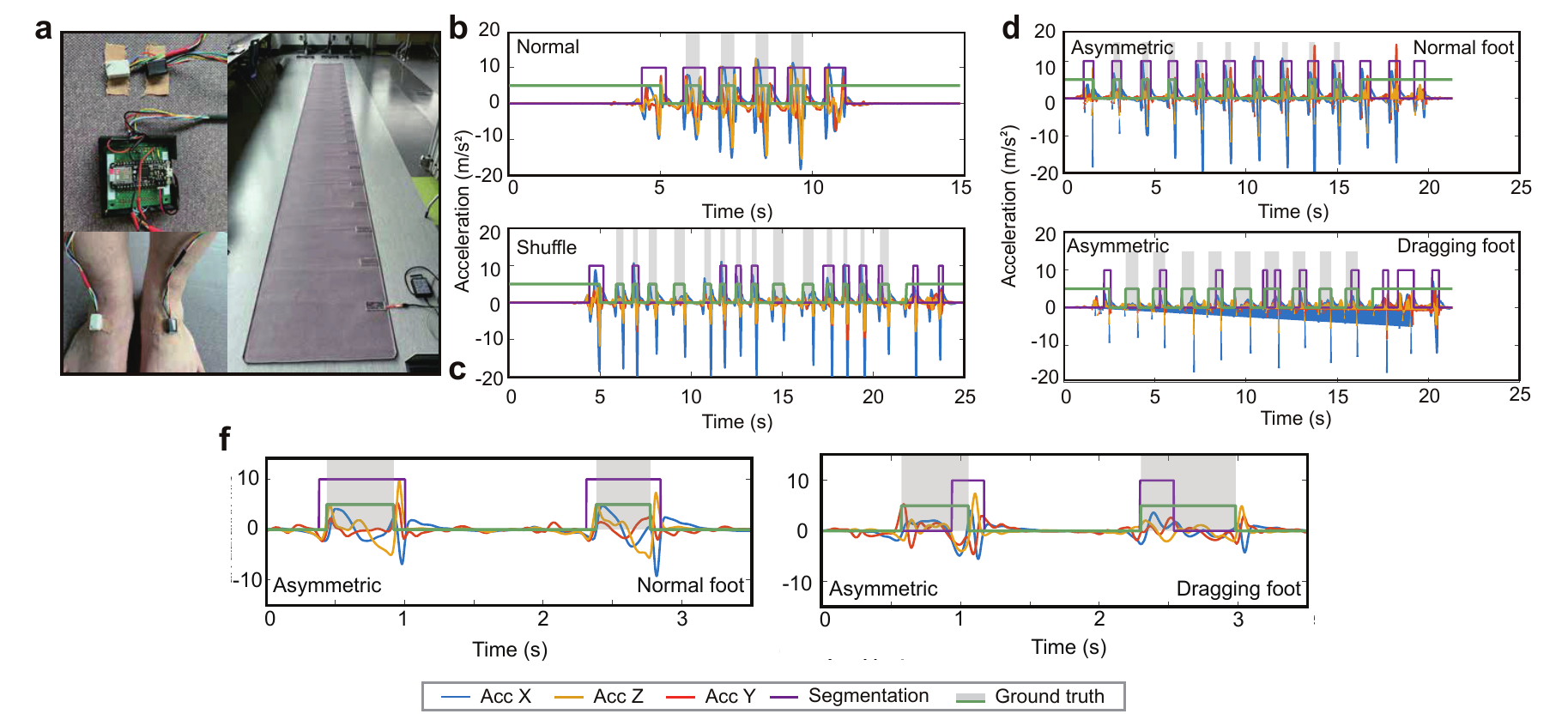}}
  \caption{(a). Experimental setup for data collection: IMUs are attached to the feet, and data is recorded while participants walk over a 1-meter-long Gaitrite mat. (b)-(d). Data type with different walking patterns and ground truth values after sensor fusion and gravity subtraction. The purple lines are the ZUPT segmentation results with missed and overcounted steps. (b). Normal walking usually displays consistent stride length and fixed cadence. (c). Parkinson's gaits have alternating small and shuffling gaits. (d). Stroke gaits exhibit asymmetric gait status where the normal foot has stronger power in moving control, while less motion is observed in the dragging foot over the accelerometer and gyroscope. (f). A detailed comparison zooming in on both normal and dragging feet. }
  \label{fig:setup}
\end{figure*}

\subsection{Ethical Considerations}
\label{sec:ethical}
The fieldwork reported in this paper was covered by UBC Ethics Certificate number H21-02052. Wearable sensors and data collection for healthcare applications raise ethical concerns that demand careful consideration. Our study took measures to address these concerns. Participants were fully informed of the study's purpose, potential risks, and benefits, and provided informed consent. Personal information was kept confidential, and data were used exclusively for research purposes. Our wearable sensors were designed to be safe, minimizing discomfort and skin irritation. We monitored participants to prevent extended use of sensors that could cause skin irritation. We also complied with data protection laws by securely storing all collected data and taking measures to prevent unauthorized access. Our study was conducted ethically and responsibly. 

\section{Results and Discussion}
\begin{figure*}[htbp]
\centering
{\includegraphics[width=0.9\linewidth]{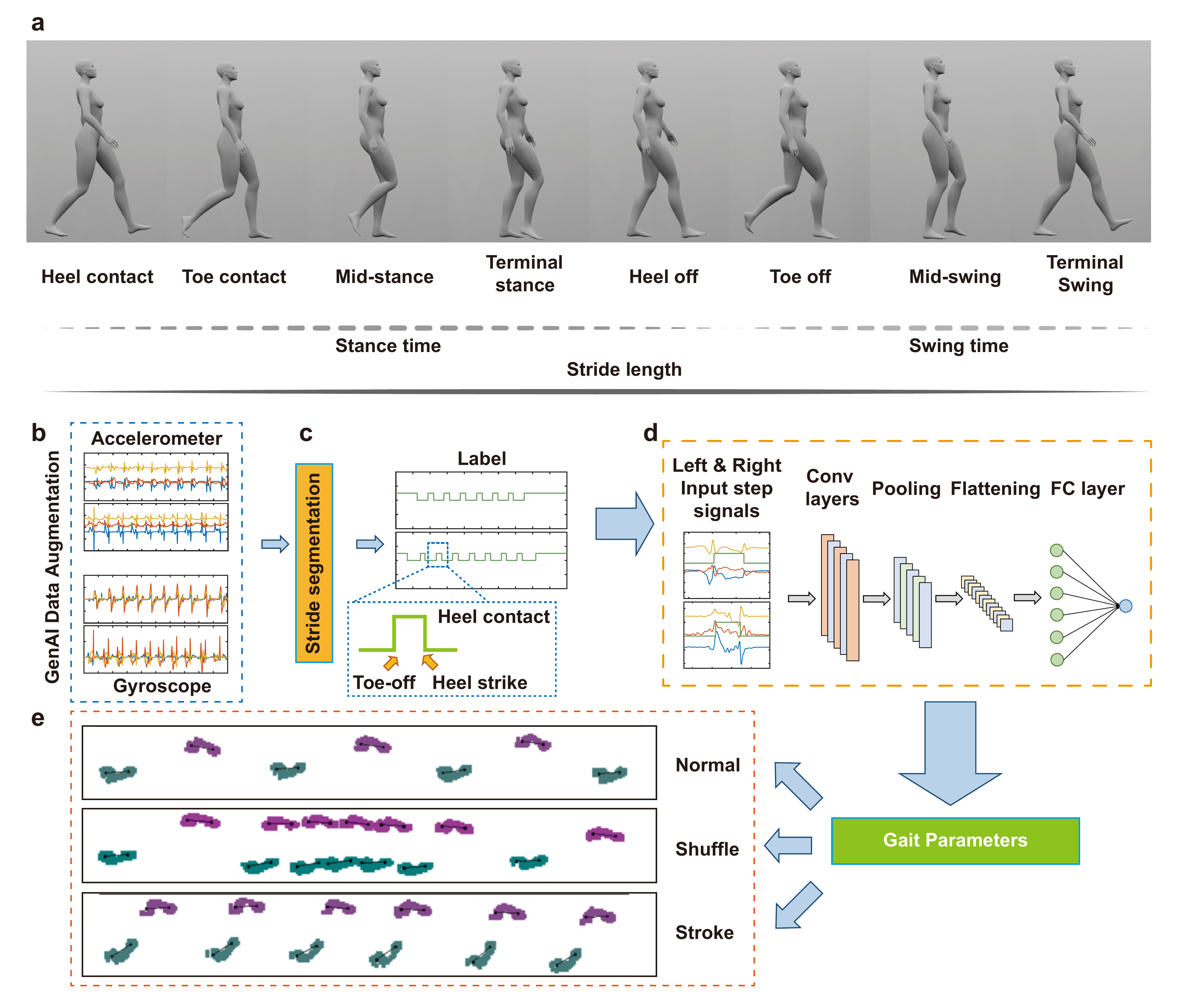}}
  \caption{Schematic process used by GaitMotion for multiple tasks learning. (a) The walking process is segmented into gait cycles, encompassing both a stance phase and a swing phase. For stride segmentation, the emphasis is placed on determining the heel strike and toe-off events, which are pivotal points in all patterns of gaits. (b) The segmentation process takes accelerometer and gyroscope data and learns to identify the events as shown in (c). Stance time is the time between the heel strike and toe-off, and swing time is the time between toe-off and the next heel strike. (d) Gait parameters such as stride length are calculated using a CNN network that takes in raw accelerometer and gyroscope data for each step, as illustrated in (e).}
  \label{fig:gaitcycle}
\end{figure*}

Our study leverages an extensive dataset that encapsulates a wide range of gait parameters across different tasks. This approach allows for a more nuanced understanding of gait dynamics, as illustrated in \autoref{fig:gaitcycle}. The gait cycle, depicted in \autoref{fig:gaitcycle}a, encompasses multiple critical states, including the stance and swing phases, alongside the HS and TO events. \autoref{fig:gaitcycle}b-e further delineates the corresponding IMU signals generated during the walking cycle, emphasizing the HS and TO events. The timing of these events, coupled with detailed insights into the stance and swing phases, including the duration of single and double support time, is pivotal for assessing locomotion stability and balance \cite{SILVA2020225}. These gait cycle parameters play an important role in the variability observed during the walking phase, which can be a marker for various health conditions, such as joint pathology, diminished muscle strength, and restricted range of motion \cite{hausdorff2009gait, hausdorff2005gait}. Understanding these components is essential for healthcare professionals in effectively monitoring and treating gait disorders.

The details of the data acquisition process and IMU walking patterns are present in the \autoref{fig:setup} and the Method section. We provide data for multiple-task learning, which includes walking type classification (normal, shuffle, or stroke pattern in  \autoref{fig:setup} b-d), step segmentation (\autoref{fig:setup} f), and gait parameter estimation. This is coupled with symmetry analysis for pathological gait patients, gait cycle estimation, and gait variability assessment. We collect data from IMUs attached to feet \cite{bamberg2008gait,teufl2018towards}, which require less signal pre-processing and acquire the most direct information on the foot movement and ground contact compared to IMUs attached to the shank or thigh \cite{panebianco2018analysis}.  The ground truth data is captured using the Gaitrite system \cite{bilney2003concurrent} in clinical settings for measuring gait parameters.  

\begin{figure*}[t]
\centering
{\includegraphics[width=0.9\linewidth]{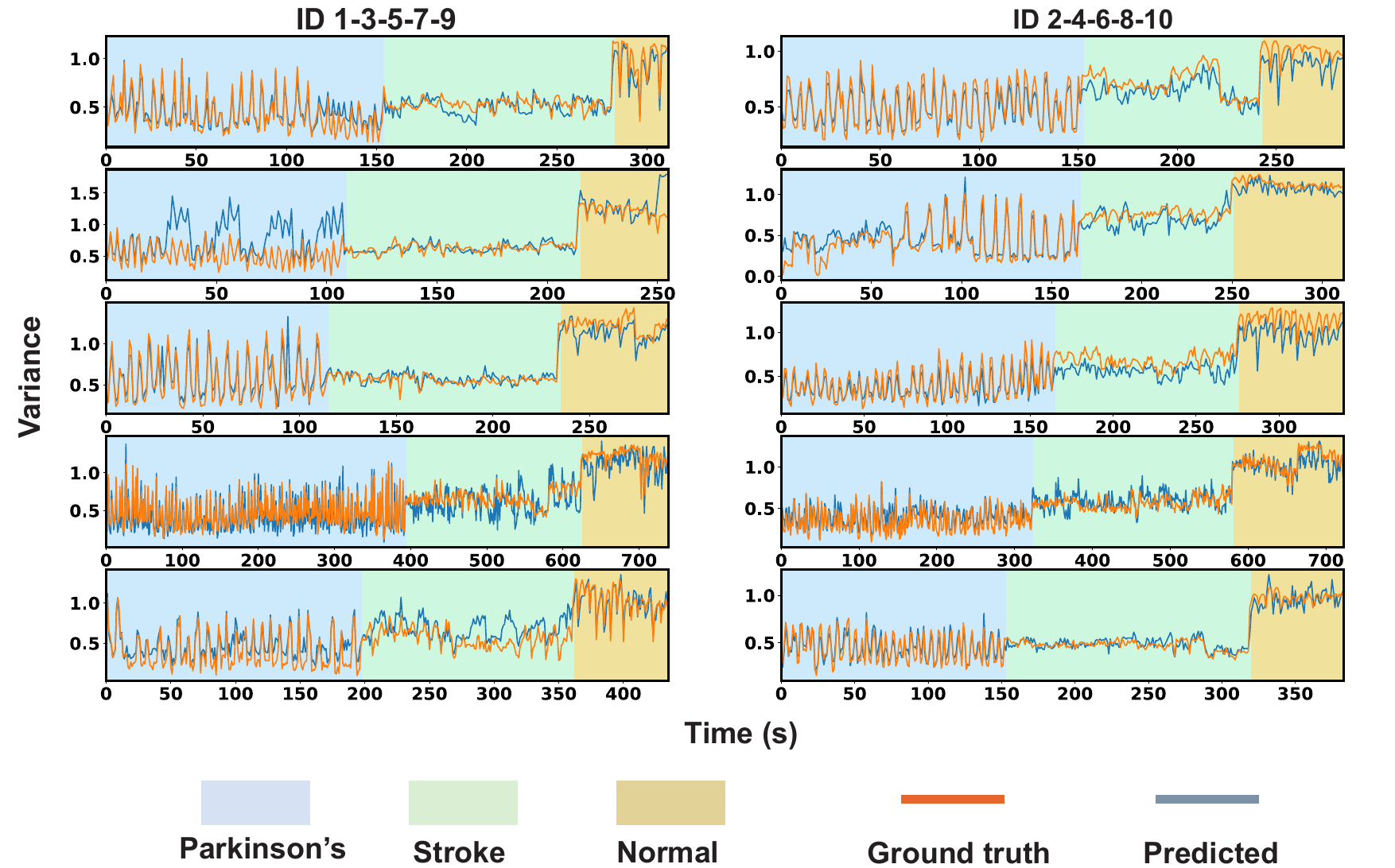}}
  \caption{Comparison of stride length estimation and ground truth with subject ID 1-10. The results include performance for Parkinson's, Stroke, and Normal gaits separately.}
  \label{fig:prediction}
  
\end{figure*}

\subsection{GaitMotion Analysis}

In this study, step segmentation is conducted using ground truth data obtained from the Gaitrite system. Typically, this process is performed by ZUPT\cite{stirling2003innovative}. However, ZUPT often encounters challenges such as missing or over-counting steps, particularly in pathological gait patterns. Accurate step segmentation is crucial for reliable gait parameter estimation, as inaccuracies in step counting can lead to significant errors. To address these limitations, the semantic segmentation task is a more robust approach. We directly employ ground truth values from Gaitrite for enhanced precision in step segmentation. 
The dataset was partitioned based on subject IDs to account for the unique behavioral variations inherent in each individual, which can lead to domain shifts. Our training protocol encompassed gait data from nine subjects, representing normal, shuffle, and stroke-related gait patterns. The remaining subjects' walking data were reserved for testing. This approach was designed to simulate scenarios where the model encounters data from previously unseen participants, thereby evaluating the model's ability to maintain accurate gait parameter estimation under such conditions. 

\begin{figure*}[t]
\centering
 {\includegraphics[width=0.8\linewidth]{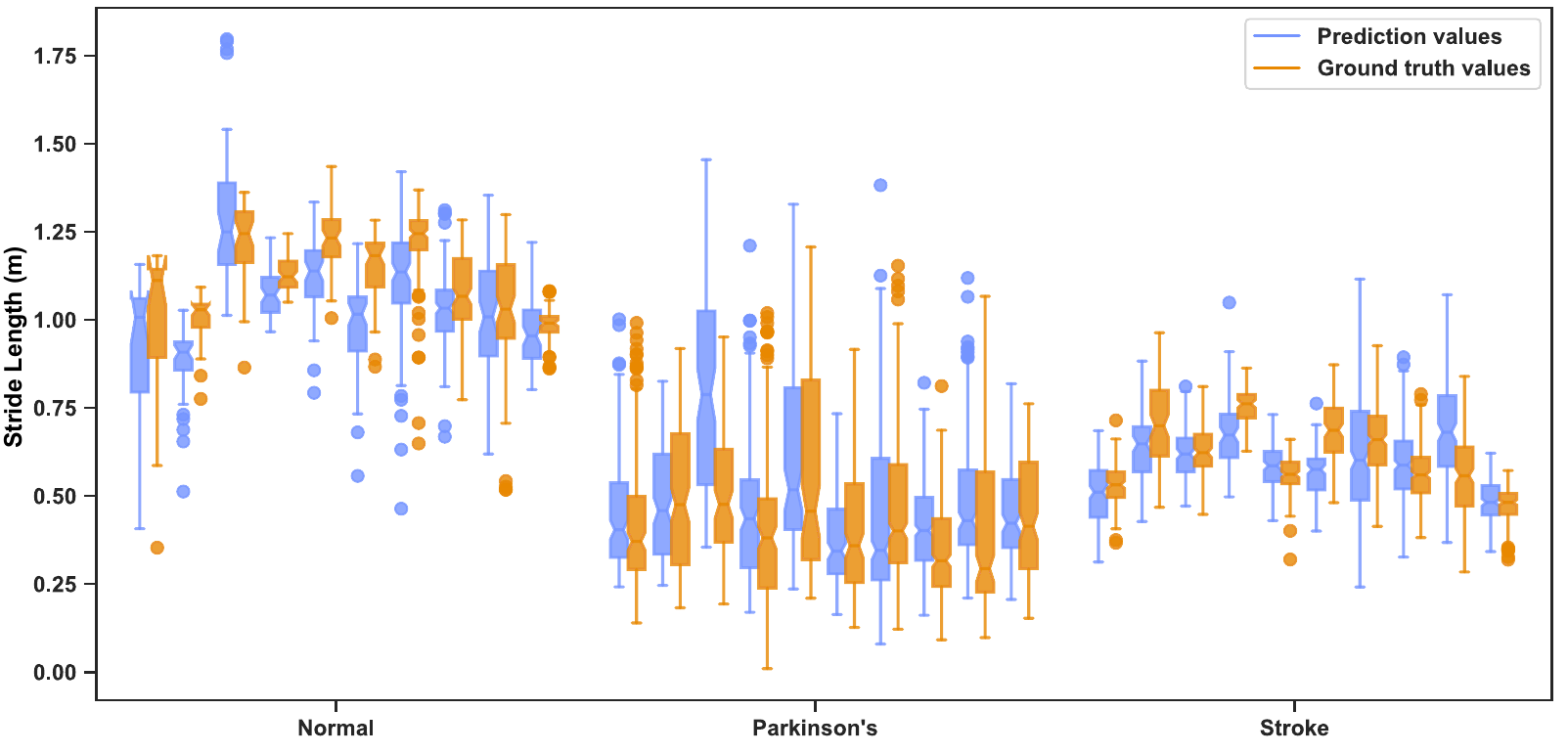}}
  \caption{Boxplot errors to separately present the summary of stride length in both pathological (Parkinson's and Stroke) and normal gaits for individual participants for comparison. }
  \label{fig:prediction_boxplot}
 
\end{figure*}

In order to address the domain shift observed among individual participants, we have illustrated the estimation results corresponding to each subject ID in \autoref{fig:prediction}. This visualization clearly delineates the distinctions between normal and abnormal walking patterns. Typically, a normal gait is characterized by larger stride lengths in comparison to those observed in Parkinson's and stroke patients. Shuffle gait patterns are the most variable, displaying a combination of shorter strides interspersed with occasional larger steps. This pattern aligns with the characteristic freezing and irregular step lengths commonly observed in Parkinson's disease patients. The estimation results, including outliers, are visually represented through a boxplot in \autoref{fig:prediction_boxplot}. Outliers are present in both normal and pathological gait patterns. Specifically, in normal gaits, the stride lengths vary from 0.586 to 1.435 meters. In contrast, stride lengths for stroke and Parkinson's conditions demonstrate a range from 0.284 to 0.963 meters and 0.0098 to 1.207 meters, respectively.

\begin{table*}[t]
\small
\centering
\caption{Statistical outcomes for each walking pattern and individual. RMSE was employed to gauge the accuracy of stride length estimations and their deviation from the ground truth. The mean value, accompanied by the STD, serves to quantify the variance in stride length for each individual and each walking pattern.}
\label{tab:RMSE_sum}
{\begin{tabular}{|c|cccc|ccc|}
\hline
 & \multicolumn{4}{c|}{RMSE (meter)} & \multicolumn{3}{c|}{Mean value (meter)} \\ \hline
subject ID & \multicolumn{1}{c|}{Normal} & \multicolumn{1}{c|}{Shuffle} & \multicolumn{1}{c|}{Stroke} & All & \multicolumn{1}{c|}{Normal} & \multicolumn{1}{c|}{Shuffle} & Stroke \\ \hline
1 & \multicolumn{1}{c|}{0.109} & \multicolumn{1}{c|}{0.091} & \multicolumn{1}{c|}{0.085} & 0.091 & \multicolumn{1}{c|}{1.003$\pm$0.205} & \multicolumn{1}{c|}{0.43$\pm$0.2} & 0.53$\pm$0.058 \\ \hline
2 & \multicolumn{1}{c|}{0.154} & \multicolumn{1}{c|}{0.067} & \multicolumn{1}{c|}{0.107} & 0.098 & \multicolumn{1}{c|}{1.005$\pm$0.066} & \multicolumn{1}{c|}{0.498$\pm$0.201} & 0.716$\pm$0.12 \\ \hline
3 & \multicolumn{1}{c|}{0.239} & \multicolumn{1}{c|}{0.373} & \multicolumn{1}{c|}{0.057} & 0.263 & \multicolumn{1}{c|}{1.217$\pm$0.107} & \multicolumn{1}{c|}{0.51$\pm$0.178} & 0.622$\pm$0.071 \\ \hline
4 & \multicolumn{1}{c|}{0.081} & \multicolumn{1}{c|}{0.098} & \multicolumn{1}{c|}{0.124} & 0.102 & \multicolumn{1}{c|}{1.133$\pm$0.049} & \multicolumn{1}{c|}{0.409$\pm$0.226} & 0.756$\pm$0.055 \\ \hline
5 & \multicolumn{1}{c|}{0.135} & \multicolumn{1}{c|}{0.136} & \multicolumn{1}{c|}{0.057} & 0.111 & \multicolumn{1}{c|}{1.225$\pm$0.095} & \multicolumn{1}{c|}{0.567$\pm$0.293} & 0.564$\pm$0.054 \\ \hline
6 & \multicolumn{1}{c|}{0.184} & \multicolumn{1}{c|}{0.082} & \multicolumn{1}{c|}{0.146} & 0.13 & \multicolumn{1}{c|}{1.147$\pm$0.095} & \multicolumn{1}{c|}{0.406$\pm$0.181} & 0.686$\pm$0.08 \\ \hline
7 & \multicolumn{1}{c|}{0.178} & \multicolumn{1}{c|}{0.116} & \multicolumn{1}{c|}{0.165} & 0.143 & \multicolumn{1}{c|}{1.22$\pm$0.114} & \multicolumn{1}{c|}{0.462$\pm$0.207} & 0.662$\pm$0.108 \\ \hline
8 & \multicolumn{1}{c|}{0.109} & \multicolumn{1}{c|}{0.09} & \multicolumn{1}{c|}{0.103} & 0.099 & \multicolumn{1}{c|}{1.077$\pm$0.114} & \multicolumn{1}{c|}{0.342$\pm$0.131} & 0.565$\pm$0.073 \\ \hline
9 & \multicolumn{1}{c|}{0.093} & \multicolumn{1}{c|}{0.137} & \multicolumn{1}{c|}{0.19} & 0.153 & \multicolumn{1}{c|}{1.019$\pm$0.182} & \multicolumn{1}{c|}{0.393$\pm$0.224} & 0.554$\pm$0.118 \\ \hline
10 & \multicolumn{1}{c|}{0.085} & \multicolumn{1}{c|}{0.075} & \multicolumn{1}{c|}{0.047} & 0.068 & \multicolumn{1}{c|}{0.983$\pm$0.05} & \multicolumn{1}{c|}{0.437$\pm$0.157} & 0.471$\pm$0.053 \\ \hline
AVG & \multicolumn{1}{c|}{0.141} & \multicolumn{1}{c|}{0.133} & \multicolumn{1}{c|}{0.122} & 0.131 & \multicolumn{1}{c|}{1.11$\pm$0.144} & \multicolumn{1}{c|}{0.433$\pm$0.207} & 0.599$\pm$0.116 \\ \hline
\end{tabular}
}
\end{table*}

The RMSE and mean value with standard deviation (STD) of the current setting are summarized in the \autoref{tab:RMSE_sum}. The mean and STD values are calculated under the scope of ground truth values. From RMSE results, we can find algorithm performance varies from subject to subject, but there is no significant discrepancy in terms of stride length estimation accuracy between pathological gaits and normal gaits after step/stride segmentation. 
Inter-subject cross-validation suggests that the domain shift among participants influences the algorithm performance to a distinctive extent. 
From the STD values and statistical data in \autoref{tab:RMSE_sum}, it's evident that while normal gaits exhibit fluctuations in stride lengths, these variations are considerably less pronounced than in the Parkinson's pattern. The stride lengths of stroke patients remain consistent, and the asymmetry between dragging feet and normal feet is better represented by the swing-to-stance phase ratio and the characteristics of IMU raw data, rather than by stride length variance. 
Variations in the gait cycle cause the swing phase duration to fluctuate, influencing the padding length for each gait. Typically, the swing stance in both Stroke and Parkinson's patients is shorter, necessitating an extended padding period. Selecting an optimal step/stride length can yield improved results. In our algorithm, we've set a stride length of 800 sample points as a constant input to ensure all swing phases within each gait cycle are captured. A stride length that's too short might not capture all changes in the gyroscope and accelerometer during the swing phase. Conversely, an overly extended stride length can lead to a ratio where effective movement is overshadowed by the stationary period, making it challenging for the network to discern crucial features effectively. 

\begin{figure*}[t]
\centering
{\includegraphics[width=0.9\linewidth]{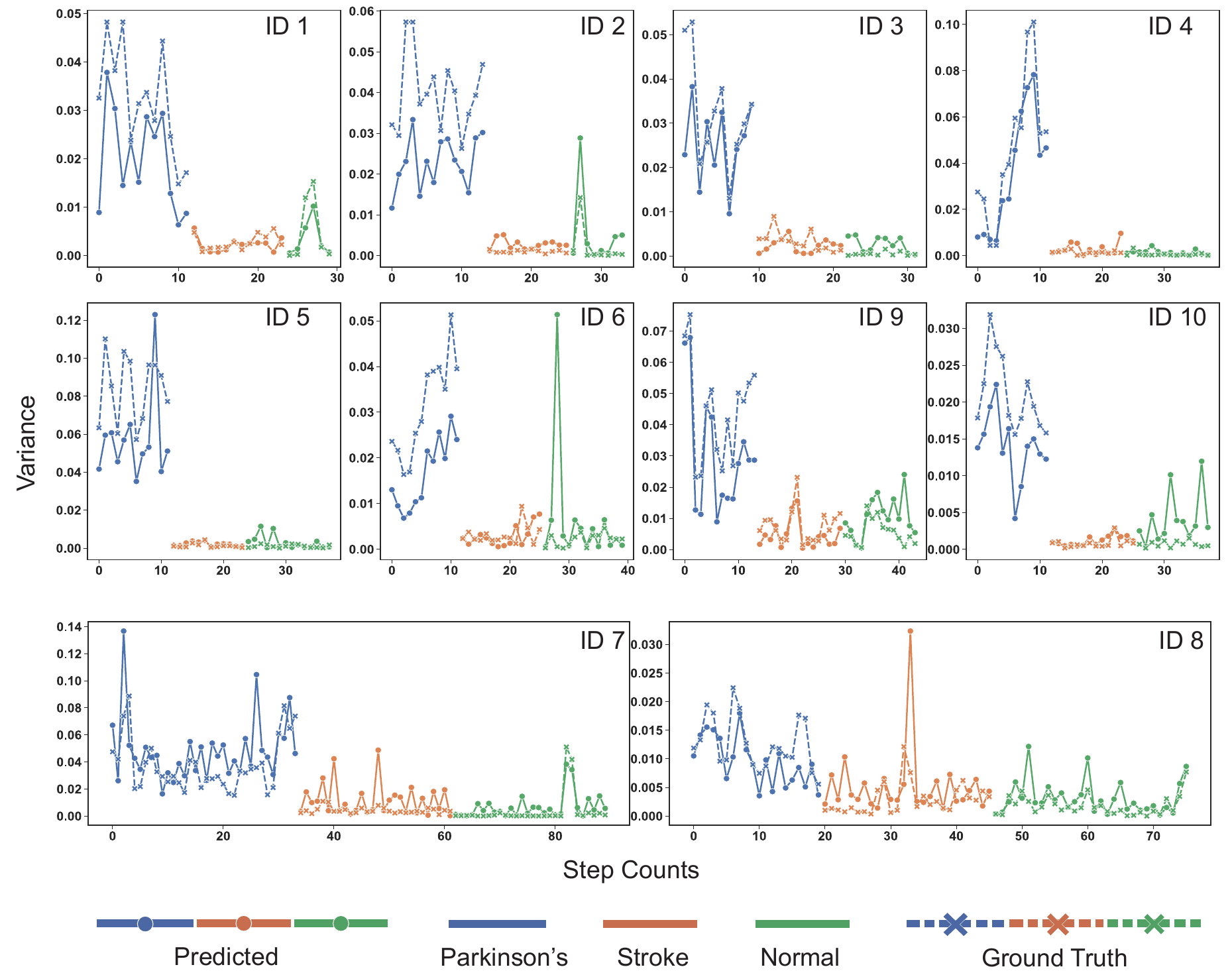}}
  \caption{Comparison of stride length variance and ground truth calculated from Gaitrite system with subject ID 1-10. The stride-to-stride gait variance results are obvious that Parkinson's gaits have the most unstable while normal and Stroke gaits have comparable levels.}
  \label{fig:stride_variance}
  
\end{figure*}

Beyond RMSE, stride-by-stride variance serves as a crucial indicator of disease severity. As depicted in \autoref{fig:stride_variance}, Parkinson's patients exhibit the highest variance, whereas the gaits of normal and stroke patients tend to be more consistent. 
Comparative analyses between conventional biomechanical algorithms (ZUPT) and ML performance are detailed in the  \ref{apd:CNN}. Our findings indicate a higher accuracy and generalizability in predicting gait parameters for both pathological and healthy individuals using our approach.

\subsection{Transfer Learning Validate} 
We aim to bridge the gap between limited datasets with labels from actual patients and the critical importance of such labeled data. We demonstrate that our dataset offers great value for the real clinical population, especially when they are unable to provide high-quality labels due to the constraints of the current Gaitrite. To validate the applicability of our dataset to clinical populations, we evaluated our model using real patient data from eGait \cite{rampp2014inertial, barth2015stride} (age 82.1 ± 6.5 years). In our transfer learning approach, we: 1). Pre-trained the model using our dataset. 2). Loaded the parameters from this pre-trained model and kept them fixed. 3). Fine-tuned only the last fully connected layers using data from four real patients from eGait, with approximately 20 steps available for training and around 1,200 steps for evaluation. The RMSE present in our transfer learning method is 0.201~m. More detailed results are attached in \autoref{tbl:transfer} for reference. Given the limited training labels, the RMSE reflects a strong performance by our model which underscores the potential of the GaitMotion dataset as a robust foundation for pre-training models designed for real patients with gait disorders. By leveraging this pre-training, we require only a minimal number of labels to fine-tune the model parameters. This approach facilitates accurate gait parameter estimations without the need to collect extensive patient data from the Gaitrite carpet.

\begin{table}[htbp]
\centering
\caption{Statistical results of transfer learning results in the unit of cm for Mean, MAE, and RMSE. STD: Standard Deviation. MAE: Mean Absolute Error. MSE: Mean Squared Error. RMSE: Root Mean Squared Error.}
\label{tbl:transfer}

\begin{tabular}{|c|cc|}
\hline
Matrix                & \multicolumn{2}{c|}{Value}                      \\ \hline
\multirow{2}{*}{Mean ($m$)} & \multicolumn{1}{c|}{Predictions} & Ground Truth \\ 
                      & \multicolumn{1}{c|}{0.8175}      & 0.7973       \\ \hline
STD ($m$)             & \multicolumn{1}{c|}{0.1880}      & 0.2329       \\ \hline
MAE ($m$)               & \multicolumn{2}{c|}{0.1589}                     \\ \hline
MSE ($m^{2}$)                & \multicolumn{2}{c|}{0.0405}                     \\ \hline
RMSE ($m$)               & \multicolumn{2}{c|}{0.2012}                     \\ \hline
\end{tabular}
\end{table}

\begin{figure*}[htbp]
\centering
  {\includegraphics[width=0.8\linewidth]{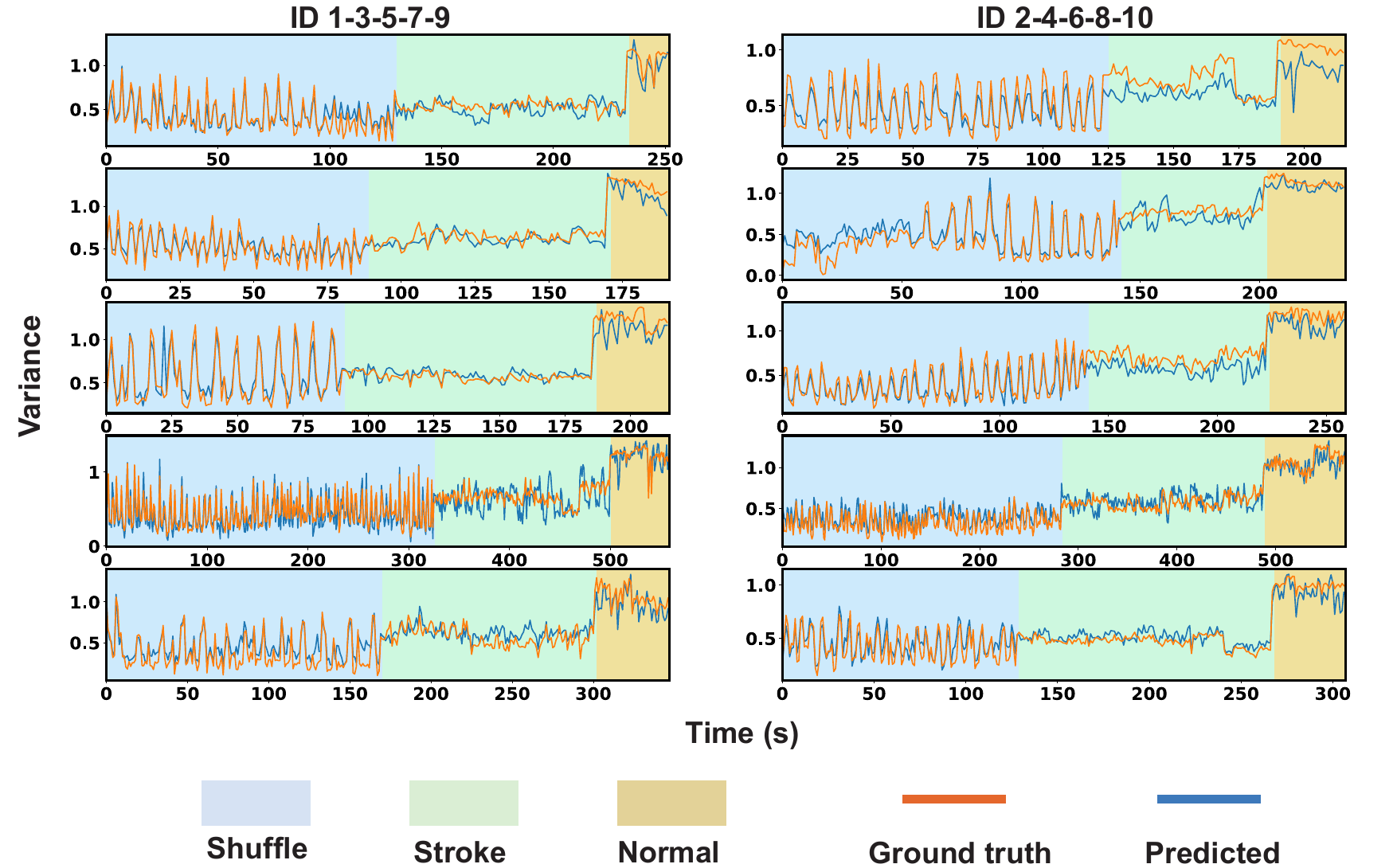}}
  \caption{Comparison of stride length estimation and ground truth with subject ID 1-10 after discarding the first and last steps.}

  \label{fig:figure1_discard_ai}
\end{figure*}

\begin{figure*}[htbp]
\centering
  {\includegraphics[width=0.9\linewidth]{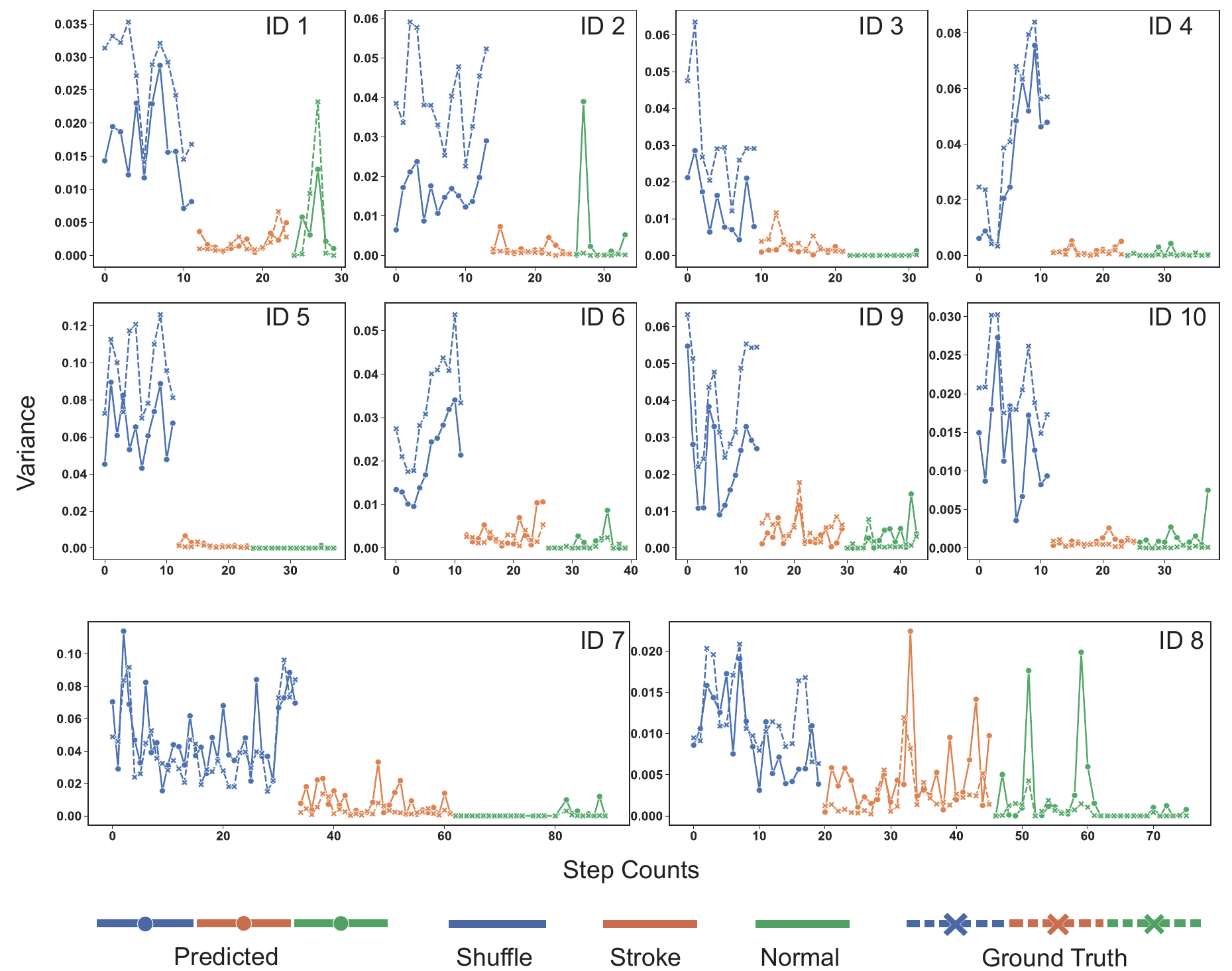}}
  \caption{Comparison of stride length variance and ground truth with subject ID 1-10 after discarding the first and last steps.}

  \label{fig:figure2_discard_ai}
\end{figure*}

\begin{figure*}[htbp]
\centering
{\includegraphics[width=0.7\linewidth]{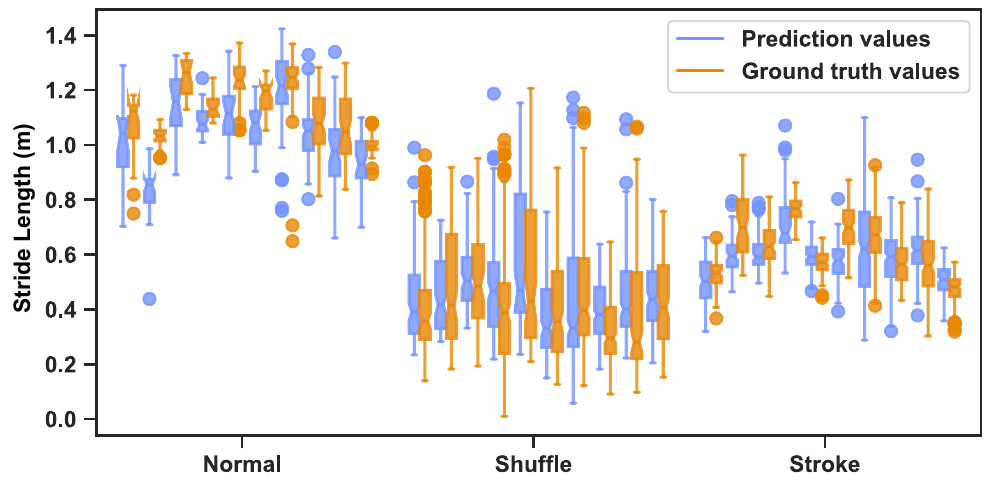}}
  \caption{Individual participant’s stride length summary in pathological and normal gaits after discarding the first and last step, visualized in the boxplot error. }
  
  \label{fig:boxplot_discard}
\end{figure*}

\subsection{Gait Parameter Prediction Outcomes}\label{apd:CNN}

\paragraph{Asymmetry and Outlier Analysis}
\label{sec:asy_outlier}
The patterns for the normal foot and dragging foot are obviously different. The normal foot displays better movement in both acceleration and gyroscope value, while the dragging foot has a lower range for both sensors. This can be used as the gait pattern feature to distinguish the gaits between normal people and Stroke patients. Statistical results comparison before and after step discard, including the stride boundary results for the ground truth values, and the outlier in each walking type for both prediction and ground truth values are summarized in the \autoref{tab:sta_sum}. The outliers decreased after discarding the first and last steps. However, for normal walking, limited to the carpet length, we only have four to six steps for each trial. After leaving out the first and last steps, the total number of strides decreases drastically. Future methods will be recommended to include the first and last steps but with separate condition statements to discard the influenced steps. In normal gaits, the stride length ranges from 0.586 to 1.435 meters, while the Stroke and Parkinson's stride length fluctuates between 0.284 to 0.963 and 0.0098 to 1.207 separately. The outliers for normal, Parkinson's, and Stroke gaits are 24/29 (algorithm predictions/ground truth values), 21/29, and 5/14 over a total number of 683, 1921, and 1449 strides. After simply discarding the first and last step of all walking trials to avoid possible influence, the normal, Stroke, and Parkinson's stride length boundaries are 0.814/1.372, 0.303/0.963, and 0.0098/1.207 separately. And the outliers for each walking pattern (normal, Parkinson's, and Stroke) are corrected to 10/16 (algorithm predictions/ground truth values), 12/28, and 15/13 over a total number of 383, 1617, and 1133 steps. 



\begin{table*}[htbp]
\small
\centering
\caption{Statistical results comparison before and after step discard. The stride boundary results are summarized for the ground truth values. }
\label{tab:sta_sum}
{\resizebox{0.85\linewidth}{!}{ 
\begin{tabular}{|c|ccccc|ccccc|}
\hline
Type & \multicolumn{5}{c|}{No Stride Discarded} & \multicolumn{5}{c|}{Discard Frist and Last Strides} \\ \hline
\multirow{2}{*}{Data} & \multicolumn{2}{c|}{Stride Boundary} & \multicolumn{2}{c|}{Outlier} & \multirow{2}{*}{\begin{tabular}[c]{@{}c@{}}Total\\ Stride\end{tabular}} & \multicolumn{2}{c|}{Stride Boundary} & \multicolumn{2}{c|}{Outlier} & \multirow{2}{*}{\begin{tabular}[c]{@{}c@{}}Total\\ Stride\end{tabular}} \\ 
 & \multicolumn{1}{c|}{Min} & \multicolumn{1}{c|}{Max} & \multicolumn{1}{c|}{\begin{tabular}[c]{@{}c@{}}Prediction\\ Values\end{tabular}} & \multicolumn{1}{c|}{\begin{tabular}[c]{@{}c@{}}Ground\\ Truth\end{tabular}} &  & \multicolumn{1}{c|}{Min} & \multicolumn{1}{c|}{Max} & \multicolumn{1}{c|}{\begin{tabular}[c]{@{}c@{}}Prediction\\ Values\end{tabular}} & \multicolumn{1}{c|}{\begin{tabular}[c]{@{}c@{}}Ground\\ Truth\end{tabular}} &  \\ \hline
Normal & \multicolumn{1}{c|}{0.568} & \multicolumn{1}{c|}{1.435} & \multicolumn{1}{c|}{24} & \multicolumn{1}{c|}{29} & 683 & \multicolumn{1}{c|}{0.814} & \multicolumn{1}{c|}{1.372} & \multicolumn{1}{c|}{10} & \multicolumn{1}{c|}{16} & 383 \\ \hline
Parkinson & \multicolumn{1}{c|}{0.0098} & \multicolumn{1}{c|}{1.207} & \multicolumn{1}{c|}{21} & \multicolumn{1}{c|}{29} & 1921 & \multicolumn{1}{c|}{0.0098} & \multicolumn{1}{c|}{1.207} & \multicolumn{1}{c|}{12} & \multicolumn{1}{c|}{18} & 1617 \\ \hline
Stroke & \multicolumn{1}{c|}{0.284} & \multicolumn{1}{c|}{0.963} & \multicolumn{1}{c|}{5} & \multicolumn{1}{c|}{14} & 1449 & \multicolumn{1}{c|}{0.303} & \multicolumn{1}{c|}{0.963} & \multicolumn{1}{c|}{15} & \multicolumn{1}{c|}{13} & 1133 \\ \hline
\end{tabular}}}
\end{table*}

The visualized results with the outliers are displayed in the boxplot \autoref{fig:prediction_boxplot}. Most of the outliers in normal gaits is at the minimum boundary, which may be caused by the first step and last step on the carpet. Since the Gaitrite carpet is at a fixed length, some participants' walking patterns might be influenced and unconsciously shorten their step when the carpet ends in the middle of one step. Future processing will be recommended to carefully recognize if the first and last steps in each trial are impacted and should be discarded or not. Meanwhile, the outliers in the Stroke and Parkinson's walking pattern are generally at the boundary of maximum stride length. This agrees with the irregular steps in pathological gaits. Compared to the prediction results for per-subject after discarding the first and last steps in \autoref{fig:figure1_discard_ai}, most subjects are not influenced, except for the third subject, where the prediction values are much better after discarding. The corresponding variance for each subject is illustrated in the \autoref{fig:figure2_discard_ai}. The variance is decreased for most of the subjects (subject with ID 1, 4, 7, 8, 9) while some still remain the same. The stride length summary of each participant was analyzed in \autoref{fig:boxplot_discard} for both pathological and normal gaits, after excluding the first and last steps. 

The RMSE and mean-STD values for discarding the first and last steps are displayed in the \autoref{tab:RMSE_mean_discard}. After comparison with \autoref{tab:RMSE_sum}, it's easy to find out that the RMSE values are not affected by those influenced steps. The prediction outcome is determined only by the recording signals of the accelerometer and gyroscope. That's also the reason why the LSTM has similar outcomes for normal and pathological gaits. But the affected steps are indeed impacting the STD of gait indicators for the disease diagnosis of patients because they are disturbing the gait parameter results.

\begin{table*}[htb]
\small
\centering
\caption{Statistical results for each walking pattern and subject after discarding the first and last steps. The RMSE is defined to measure the prediction accuracy and deviation of the ground truth. The mean value with standard deviation aims to quantify the stride length variance within each subject. }
\label{tab:RMSE_mean_discard}
{\resizebox{0.85\linewidth}{!}{ 
\begin{tabular}{|c|cccc|ccc|}
\hline
 & \multicolumn{4}{c|}{RMSE ($m$)} & \multicolumn{3}{c|}{Mean value ($m$)} \\ \hline
subject ID & \multicolumn{1}{c|}{Normal} & \multicolumn{1}{c|}{Shuffle} & \multicolumn{1}{c|}{Stroke} & All & \multicolumn{1}{c|}{Normal} & \multicolumn{1}{c|}{Shuffle} & Stroke \\ \hline
1 & \multicolumn{1}{c|}{0.115} & \multicolumn{1}{c|}{0.089} & \multicolumn{1}{c|}{0.088} & 0.090 & \multicolumn{1}{c|}{1.066$\pm$0.125} & \multicolumn{1}{c|}{0.416$\pm$0.190} & 0.532$\pm$0.053 \\ \hline
2 & \multicolumn{1}{c|}{0.225} & \multicolumn{1}{c|}{0.095} & \multicolumn{1}{c|}{0.143} & 0.132 & \multicolumn{1}{c|}{1.029$\pm$0.038} & \multicolumn{1}{c|}{0.477$\pm$0.206} & 0.716$\pm$0.121 \\ \hline
3 & \multicolumn{1}{c|}{0.133} & \multicolumn{1}{c|}{0.082} & \multicolumn{1}{c|}{0.069} & 0.084 & \multicolumn{1}{c|}{1.248$\pm$0.066} & \multicolumn{1}{c|}{0.512$\pm$0.183} & 0.629$\pm$0.072 \\ \hline
4 & \multicolumn{1}{c|}{0.077} & \multicolumn{1}{c|}{0.114} & \multicolumn{1}{c|}{0.133} & 0.114 & \multicolumn{1}{c|}{1.141$\pm$0.049} & \multicolumn{1}{c|}{0.412$\pm$0.225} & 0.770$\pm$0.047 \\ \hline
5 & \multicolumn{1}{c|}{0.165} & \multicolumn{1}{c|}{0.131} & \multicolumn{1}{c|}{0.051} & 0.110 & \multicolumn{1}{c|}{1.233$\pm$0.087} & \multicolumn{1}{c|}{0.550$\pm$0.305} & 0.573$\pm$0.047 \\ \hline
6 & \multicolumn{1}{c|}{0.127} & \multicolumn{1}{c|}{0.076} & \multicolumn{1}{c|}{0.145} & 0.110 & \multicolumn{1}{c|}{1.180$\pm$0.058} & \multicolumn{1}{c|}{0.403$\pm$0.190} & 0.701$\pm$0.077 \\ \hline
7 & \multicolumn{1}{c|}{0.127} & \multicolumn{1}{c|}{0.104} & \multicolumn{1}{c|}{0.170} & 0.130 & \multicolumn{1}{c|}{1.229$\pm$0.120} & \multicolumn{1}{c|}{0.455$\pm$0.205} & 0.669$\pm$0.110 \\ \hline
8 & \multicolumn{1}{c|}{0.093} & \multicolumn{1}{c|}{0.100} & \multicolumn{1}{c|}{0.100} & 0.099 & \multicolumn{1}{c|}{1.085$\pm$0.109} & \multicolumn{1}{c|}{0.322$\pm$0.117} & 0.571$\pm$0.077 \\ \hline
9 & \multicolumn{1}{c|}{0.138} & \multicolumn{1}{c|}{0.132} & \multicolumn{1}{c|}{0.114} & 0.127 & \multicolumn{1}{c|}{1.066$\pm$0.124} & \multicolumn{1}{c|}{0.374$\pm$0.216} & 0.571$\pm$0.109 \\ \hline
10 & \multicolumn{1}{c|}{0.106} & \multicolumn{1}{c|}{0.088} & \multicolumn{1}{c|}{0.056} & 0.078 & \multicolumn{1}{c|}{1.000$\pm$0.039} & \multicolumn{1}{c|}{0.424$\pm$0.150} & 0.471$\pm$0.055 \\ \hline
AVG & \multicolumn{1}{c|}{0.141} & \multicolumn{1}{c|}{0.134} & \multicolumn{1}{c|}{0.122} & 0.131 & \multicolumn{1}{c|}{1.110$\pm$0.144} & \multicolumn{1}{c|}{0.433$\pm$0.206} & 0.600$\pm$0.116 \\ \hline
\end{tabular}}}
\end{table*}

\paragraph{Comparison of GaitMotion and ZUPT results}
\label{apd:MLZUPTCOM}

\begin{table}[htb]
\caption{RMSE (m) for ZUPT and Machine Learning, and Average Step Length for Various Walking Types on the GaitMotion Dataset.}
\label{tbl:compare_ML_ZUPT}
 \centering\begin{tabular}{|c|c|c|c|}
\hline
 - & GaitMotion & ZUPT & Mean \\ \hline
Normal & 0.125 & 0.391 & 1.110 \\ \hline
Shuffle & 0.095 & 0.226 & 0.433 \\ \hline
Stroke & 0.102 & 0.361 & 0.599 \\ \hline
Avg & 0.103 & 0.295 & - \\ \hline
\end{tabular}
\end{table}

\begin{figure}[htbp]
 \centering
  {\includegraphics[width=1\linewidth]{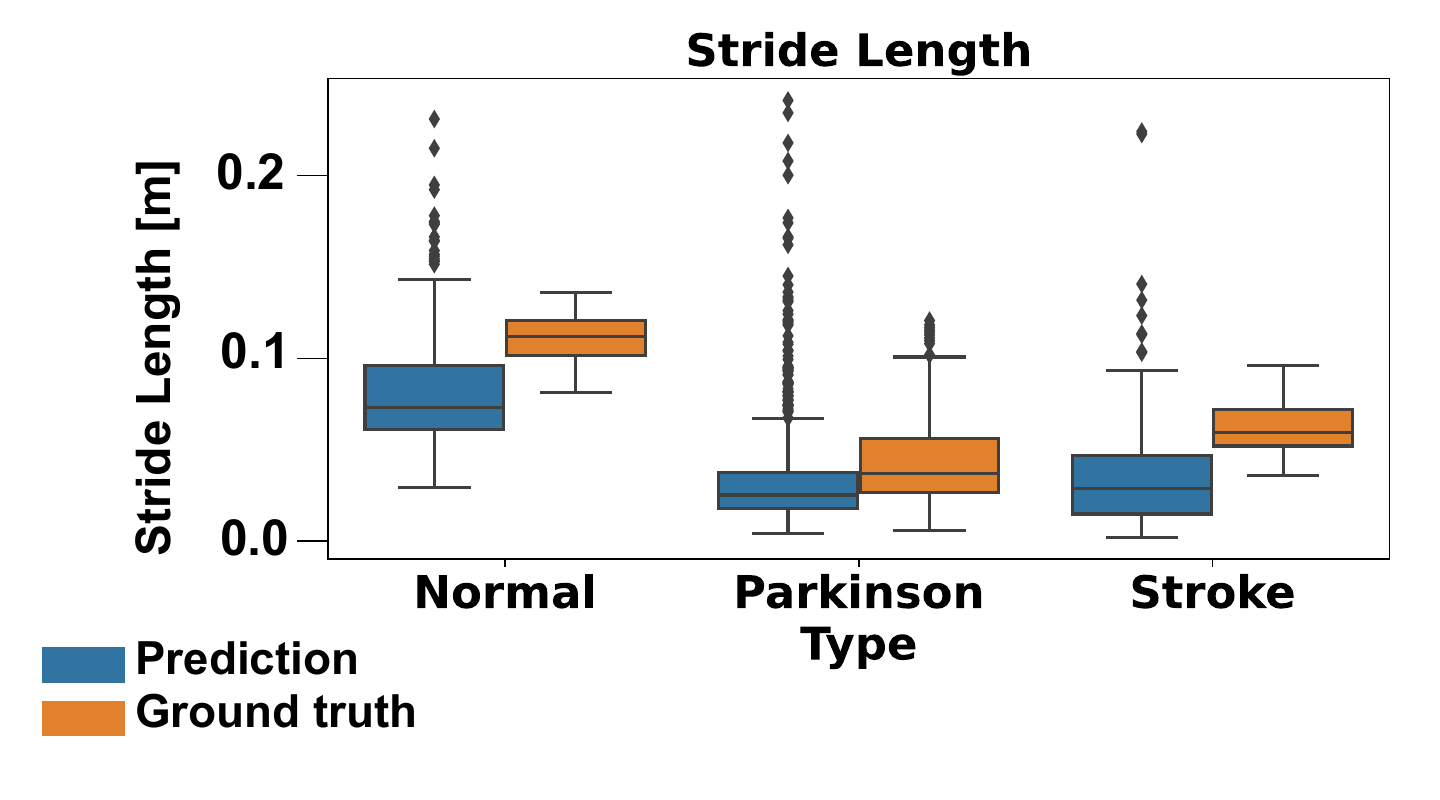}}
  \caption{Comparison of stride length prediction and ground truth values. The stride length prediction is from the ZUPT method.}
  \label{fig:stride_zupt}

\end{figure}

The RMSE error comparisons between GaitMotion and ZUPT methods are detailed in \autoref{tbl:compare_ML_ZUPT}. Given that ZUPT relies on double integration to determine stride length, it's particularly susceptible to IMU noise. Upon examining the average values across diverse walking types, it's evident that ZUPT performs optimally with healthy gait patterns. For clarity, a visualization of the stride length prediction versus the actual values can be found in \autoref{fig:stride_zupt}. Generally, the predicted values trend below the actual values and outliers manifest across all walking patterns.
 
We aim to obtain accurate gait parameter estimations from foot-worn sensors for both healthy and pathological gaits. we collected a comprehensive dataset with extensive ground truth labels from on-off ground status and spatial gait parameters, which provide valuable insights into multiple areas in healthcare applications including disease diagnosis, treatment planning, and progress monitoring. However, the dataset is generated from healthy subjects mimicking patients' walking patterns, which may suffer from domain misalignment between mimicry and truly disordered gaits. To assess the generalizability of our model in the context of real patient data, we have conducted a verification task utilizing transfer learning techniques, where we fine-tune the last fully connected layer using a limited set of labels from real patients' data in eGait \cite{rampp2014inertial}. Our model's commendable performance on the eGait suggests that the GaitMotion dataset can effectively serve as a foundation for pre-training models tailored for real patients with gait disorders. With pre-training, only minimal labels are needed to refine the model parameters, allowing for precise gait parameter estimations without the necessity to collect massive patient data from the Gaitrite carpet.

Many of the currently available datasets suffer from incomplete or absent ground truth, primarily because Gaitrite carpets are sensitive to adhesive footsteps and walking-assist devices which are essential for patients with gait disorders. Specifically, Gaitrite may encounter difficulties in collecting data from stroke patients experiencing foot drops and Parkinson's patients experiencing rapid, small-stepped gait patterns. Ideally, Gaitrite needs distinct and alternating movements between the left and right feet. Any overlapping or adhesive footsteps can hinder accurate step segmentation and gait parameter identification.  Guiding patients to walk in the prescribed pattern can be challenging and, in certain cases, unattainable due to inherent risks and feasibility limitations. Our approach bridges this gap, enabling consistent monitoring of these patients with accurate estimations. After training on our dataset, minimal labels are needed to fine-tune the parameters. Mimicking patient gait patterns with healthy subjects can lead to a sufficient pilot study and benefit real patients.

\section{Conclusion}
With our GaitMotion framework, we tackle the challenge of insufficient wearable electronics databases and the use of few-shot machine learning methods with a GenAI data augmentation module for pathological gaits, such as those seen in Parkinson's and Stroke patients. In the realm of personalized health, GaitMotion provides a new perspective into individual gait disturbances and offers a novel approach to predicting gait patterns using a small dataset. This small dataset can be obtained without patient data, which is typically difficult to collect due to ethical and privacy considerations. GaitMotion is specifically designed to cater to patients with gait challenges, making it beneficial for clinical treatments and laboratory research. Additionally, the framework can be instrumental in other applications where distinctive gait patterns are crucial. We provide full ground truth data for multiple task learning purposes including stride segmentation and stride length estimations. Validated with data from 10 healthy subjects who mimicked patient gait patterns, GaitMotion features a few-shot machine learning algorithm that records stride length estimation errors of 0.141 meters for normal gaits, 0.133 meters for Parkinson’s, and 0.122 meters for Stroke. This approach delivers robust and consistent analyses, achieving a 65\% enhancement in overall accuracy compared to traditional methods like ZUPT.

This approach provides a platform for developing and testing new interventions and rehabilitation strategies aimed at improving gait function in individuals with Stroke and Parkinson's disease. Furthermore, with the continued development and expansion of this gait database e.g. (via federated learning across different institutions), there is the potential to uncover new insights and information about other conditions that affect gait, leading to improved diagnosis and treatment for a wider range of patients. The GaitMotion framework, based on machine learning and few-shot learning, containing pathological gait patterns in Stroke and Parkinson's disease provides a valuable opportunity for learning and further research with the exploration of a better understanding of the unique gait characteristics presented under abnormal walking conditions. 

\backmatter

\bmhead{Acknowledgements}

We express our gratitude to the volunteers who participated in the data collection experiment, as well as to the anonymous reviewers for their valuable comments and discussions. This work received partial support from Natural Sciences and Engineering Research Council (NSERC) of Canada. The opinions, findings, conclusions, and recommendations presented in this paper belong to the authors and do not necessarily represent the views of the funding agencies or the government. Special acknowledgment to Prof. Calvin Kuo for his invaluable discussions and assistance throughout the ethics application process.

\section*{Declarations}

The authors declare no competing interests.

\section*{Clinical Trail Number:}
Not applicable.

\section*{Data Availability}
The datasets analysed during the current study are available from the corresponding author on reasonable request.

\bibliography{sn-bib}

\end{document}